\documentclass[conference, 10pt]{IEEEtran}

\usepackage{cite,graphicx,amsmath,psfrag,bm}
\usepackage[usenames,dvipsnames]{xcolor}
\usepackage{mathabx}
\usepackage{subfig}
\usepackage{cancel}
\usepackage{epstopdf}
\usepackage{pstool}

\newcommand{\ve}[1]{\boldsymbol{#1}}
\newcommand{\te}[1]{\overline{\overline{#1}}}

\begin{document}

\title{Controllable Angular Scattering\\with a Bianisotropic Metasurface\vspace{-0.3cm}}

\author{\IEEEauthorblockN{Karim Achouri, and Christophe Caloz}
\IEEEauthorblockA{Dept. of Electrical Engineering, Polytechnique Montr\'{e}al,
Montr\'{e}al, QC H2T 1J3, Canada\\ Email: see http://www.calozgroup.org/}}

\maketitle

\begin{abstract}
We propose the concept of a bianisotropic metasurface with controllable angular scattering. We illustrate this concept with the synthesis and the analysis of a metasurface exhibiting controllable absorption and transmission phase as function of the incidence angle.
\end{abstract}

\IEEEpeerreviewmaketitle


\section{Introduction}

Over the past few years, metasurfaces have proven to be impressively powerful in manipulating electromagnetic waves. However, most studies have been restricted to metasurfaces performing electromagnetic transformations for a unique set of incident, reflected and transmitted waves. If the incidence angle would change, the scattered waves would experience major and uncontrollable changes compared to the specified ones. Only a few studies have attempted to analyze or synthesis metasurfaces with angle-independent scattering as, for instance, in~\cite{5272305,di2011optical,ra2015angularly}.

In this work, we propose a new technique to synthesize a metasurface with controllable angular scattering. For simplicity, we consider the case of a uniform metasurface, only transforming the phase and the amplitude of the scattered waves. The metasurface is synthesized by specifying the reflection and transmission coefficients for three different incidence angles which, by continuity, allows a relative smooth control of the angular scattering as function of the incidence angle. The synthesis of a metasurface performing three transformations requires a number of degrees of freedom which are here obtained by leveraging bianisotropy and making use of longitudinal susceptibilities~\cite{achouri2014general,7428120}.

\section{Metasurface Design}
\subsection{Metasurface Synthesis and Analysis}

A bianisotropic metasurface may be described by zero-thickness continuity conditions conventionally called GSTCs~\cite{Idemen1973,kuester2003av}. For a metasurface lying in the $xy$-plane at $z=0$, the GSTCs read
\begin{subequations}
\label{eq:gstc}
\begin{align}
\hat{z}\times\Delta \ve{H} &= j\omega P_\parallel - \hat{z}\times\nabla_\parallel M_z,\\
\Delta \ve{E}\times\hat{z} &= j\omega\mu_0 M_\parallel - \nabla_\parallel\left( \frac{P_z}{\epsilon_0}\right)\times\hat{z},
\end{align}
\end{subequations}
where $\Delta \ve{E}$ and $\Delta \ve{H}$ are the differences of the electric and magnetic fields on both sides of the metasurface and where $\ve{P}$ and $\ve{M}$ are, respectively, the electric and magnetic  polarization densities, which may be expressed in terms of bianisotropic susceptibility tensors as
\begin{subequations}
\label{eq:pola}
\begin{align}
\ve{P} &= \epsilon_0 \te{\chi}_\text{ee}\cdot\ve{E_\text{av}} + \te{\chi}_\text{em}\cdot\ve{H_\text{av}}/c_0,\\
\ve{M} &= \te{\chi}_\text{mm}\cdot\ve{H_\text{av}} + \te{\chi}_\text{me}\cdot\ve{E_\text{av}}/\eta_0,
\end{align}
\end{subequations}
where $\ve{E_\text{av}}$ and $\ve{H_\text{av}}$ are the average electric and magnetic fields on both sides of the metasurface.
\vspace{-0.2cm}
\begin{figure}[htbp]
\begin{center}
\includegraphics[width=0.5\columnwidth]{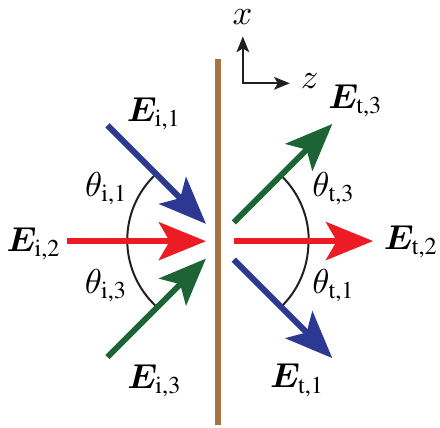}
\caption{Multiple scattering from a uniform bianisotropic metasurface.}
\label{Fig:schem}
\end{center}
\end{figure}
\vspace{-0.5cm}

Let us now consider the electromagnetic transformations depicted in Fig.~\ref{Fig:schem} where $p$-polarized incident plane waves are scattered, without rotation of polarization, by a bianisotropic metasurface. In this transformation, the only electromagnetic field components that are not zero are $E_x, E_z$ and $H_y$ and therefore only a few susceptibility components will by excited by such fields. Considering that each of the four susceptibility tensors in~\eqref{eq:pola} contains $3\times 3$ components, the only susceptibilities that are relevant to the problem of Fig.~\ref{Fig:schem} are
\begin{subequations}
\label{eq:susc}
\begin{equation}
\te{\chi}_\text{ee}=
\begin{pmatrix}
{\chi}_\text{ee}^{xx} & 0 & {\chi}_\text{ee}^{xz} \\
0 & 0 & 0 \\
{\chi}_\text{ee}^{zx} & 0 & {\chi}_\text{ee}^{zz}
\end{pmatrix},
\qquad
\te{\chi}_\text{em}=
\begin{pmatrix}
0 & {\chi}_\text{em}^{xy} & 0\\
0 & 0 & 0\\
0 & {\chi}_\text{em}^{zy} & 0\\
\end{pmatrix},
\end{equation}
\begin{equation}
\te{\chi}_\text{me}=
\begin{pmatrix}
0 & 0 & 0 \\
{\chi}_\text{me}^{yx} & 0 & {\chi}_\text{me}^{yz} \\
0 & 0 & 0
\end{pmatrix},
\qquad
\te{\chi}_\text{mm}=
\begin{pmatrix}
0 & 0 & 0\\
0 & {\chi}_\text{mm}^{yy} & 0\\
0 & 0 & 0
\end{pmatrix},
\end{equation}
\end{subequations}
where all the susceptibilities that are not excited by the fields have been set to zero for simplicity. Note that this metasurface does not induce rotation of polarization.
The susceptibility tensors in~\eqref{eq:susc} contain a total number of 9 unknown components. However, in this work, we wish to design a reciprocal metasurface, which reduces the number of unknowns to 6 since, by reciprocity, ${\chi}_\text{ee}^{xz} = {\chi}_\text{ee}^{zx}$, ${\chi}_\text{em}^{xy} = -{\chi}_\text{me}^{yx}$ and ${\chi}_\text{em}^{zy} = -{\chi}_\text{me}^{yz}$.

In order to simplify the synthesis and the analysis, we specify that the metasurface is uniform in the $xy$-plane. Then the susceptibilities are not function of $x$ and $y$ and hence the spatial derivatives on the right-hand side of~\eqref{eq:gstc} only apply to the fields and not to the susceptibilities through~\eqref{eq:pola}. This restriction means that the reflection and transmission angles follow conventional Snell's law, i.e. $\theta_\text{r} = -\theta_\text{i}$ and $\theta_\text{t} = \theta_\text{i}$.

Let us now substitute the susceptibilities~\eqref{eq:susc} into~\eqref{eq:gstc} with~\eqref{eq:pola} and enforce reciprocity. This operation reduces~\eqref{eq:gstc} to the two following equations:
\begin{subequations}
\label{eq:sys}
\begin{align}
\Delta H_y =& -j\omega\epsilon_0(\chi_\text{ee}^{xx}E_{x,\text{av}} + \chi_\text{ee}^{xz}E_{z,\text{av}}) - jk_0\chi_\text{em}^{xy}H_{y,\text{av}}\\
\begin{split}
\Delta E_x =& -j\omega\mu_0\chi_\text{mm}^{yy}H_{y,\text{av}}+jk_0(\chi_\text{em}^{xy}E_{x,\text{av}} + \chi_\text{em}^{zy}E_{z,\text{av}})\\
 &-\chi_\text{ee}^{xz} \partial_x E_{x,\text{av}} -\chi_\text{ee}^{zz} \partial_x E_{z,\text{av}} - \eta_0 \chi_\text{em}^{zy} \partial_x H_{y,\text{av}}
\end{split}
\end{align}
\end{subequations}
where $\partial_x$ is the partial derivative along $x$. The synthesis technique consists in solving~\eqref{eq:sys} for the susceptibilities in~\eqref{eq:susc}. However, as previously  mentioned, there are 6 unknown susceptibilities, and the system~\eqref{eq:sys} contains only 2 equations. This means that, to be determined, the system~\eqref{eq:sys} may be solved for three independent sets of incident, reflected and transmitted waves~\cite{achouri2014general,7428120}. Thus, the reflection and transmission coefficients of the metasurface in Fig.~\ref{Fig:schem} may be specified for three different angles of incidence. By specifying the reflection and transmission coefficients for three specific angles, one can achieve controllable quasi-continuous angular scattering since the response of the metasurface for non-specified angles de facto corresponds to an interpolation of the three specified responses.

Once the synthesis has been completed, following the aforementioned procedure, the response of the metasurface versus incidence angle for the synthesized susceptibilities may be performed by analysis, which consists in solving~\eqref{eq:sys} to determine the reflection ($R$) and transmission ($T$) coefficients.

\subsection{Illustrative Example}

We now illustrate by an example the synthesis and analysis of a bianisotropic metasurface with  controllable angular scattering. Let us consider a reflection-less transformation ($R=0$) where three incident plane waves, impinging on the metasurface at $\theta_{\text{i},1}=-45^\circ$, $\theta_{\text{i},2}=0^\circ$ and $\theta_{\text{i},3}=+45^\circ$, are transmitted with transmission coefficients $T_1 = 0.75$, $T_2 = 0.5e^{j45^\circ}$ and $T_3 = 0.25$ and transmission angles $\theta_\text{t}=\theta_\text{i}$. To synthesize the metasurface and find the corresponding susceptibilities, the electromagnetic fields, corresponding to these three transformations, are first used to define the difference and the average of the fields which are then substituted into~\eqref{eq:sys}. This leads to a system of 6 equations in 6 unknown susceptibilities, which can be easily solved. At this stage, the metasurface is synthesized, with the closed-form susceptibilities that are not shown here for the sake of conciseness.

\begin{figure}[htbp]
\begin{center}
\subfloat[]{\label{Fig:abs}
\includegraphics[width=0.7\columnwidth]{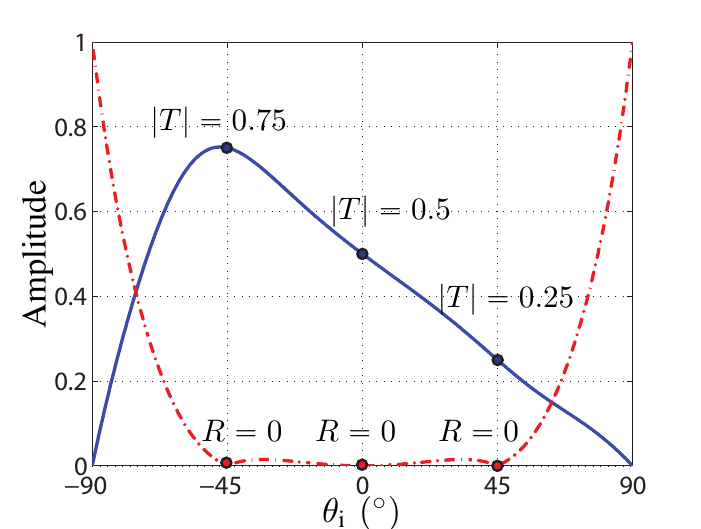}
}\\
\subfloat[]{\label{Fig:phase}
\includegraphics[width=0.7\columnwidth]{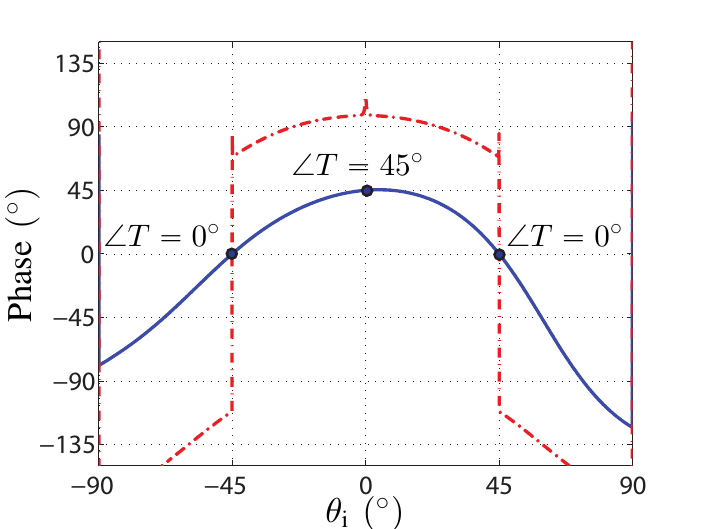}
}
\caption{Reflection (dashed red line) and transmission (solid blue line) amplitude (a) and phase (b) as function of the incidence angle for a metasurface synthesize for the transmission coefficients $T = \{0.75;0.5e^{j45^\circ};0.25\}$ (and $R=0$) at the respective incidence angles $\theta_\text{i} = \{-45^\circ;0^\circ;+45^\circ\}$.}
\label{Fig:example}
\end{center}
\end{figure}

Now, to verify that the scattered waves have the specified amplitude and phase at the three specified incidence angles and also to see the response at non-specified angles, we analyze the synthesized metasurface versus the incidence angle. For this purpose, as previously mentioned, relations~\eqref{eq:sys} are solved to determine the reflection and transmission coefficients versus $\theta_\text{i}$. The resulting amplitude and phase of the reflection and transmission coefficients are plotted in Figs.\ref{Fig:abs} and~\ref{Fig:phase}, respectively. As may be seen in these graphs, the metasurface exhibits the specified response in terms of both coefficients at the three specified angles. Moreover, the transmission exhibits a continuous amplitude decrease as $\theta_\text{i}$ increases beyond $-50^\circ$.\vspace{-0.6cm}
\vspace{-0.5cm}

\bibliographystyle{IEEEtran}
\bibliography{Achouri_aps2017}

\begin{thebibliography}{1}
\providecommand{\url}[1]{#1}
\csname url@samestyle\endcsname
\providecommand{\newblock}{\relax}
\providecommand{\bibinfo}[2]{#2}
\providecommand{\BIBentrySTDinterwordspacing}{\spaceskip=0pt\relax}
\providecommand{\BIBentryALTinterwordstretchfactor}{4}
\providecommand{\BIBentryALTinterwordspacing}{\spaceskip=\fontdimen2\font plus
\BIBentryALTinterwordstretchfactor\fontdimen3\font minus
  \fontdimen4\font\relax}
\providecommand{\BIBforeignlanguage}[2]{{%
\expandafter\ifx\csname l@#1\endcsname\relax
\typeout{** WARNING: IEEEtran.bst: No hyphenation pattern has been}%
\typeout{** loaded for the language `#1'. Using the pattern for}%
\typeout{** the default language instead.}%
\else
\language=\csname l@#1\endcsname
\fi
#2}}
\providecommand{\BIBdecl}{\relax}
\BIBdecl

\bibitem{5272305}
J.~A. Gordon, C.~L. Holloway, and A.~Dienstfrey, ``A physical explanation of
  angle-independent reflection and transmission properties of
  metafilms/metasurfaces,'' \emph{IEEE Antennas and Wireless Propagation
  Letters}, vol.~8, pp. 1127--1130, 2009.

\bibitem{di2011optical}
A.~Di~Falco, Y.~Zhao, and A.~Al{\'u}, ``Optical metasurfaces with robust
  angular response on flexible substrates,'' \emph{Applied Physics Letters},
  vol.~99, no.~16, p. 163110, 2011.

\bibitem{ra2015angularly}
Y.~Ra'di and S.~Tretyakov, ``Angularly-independent huygens' metasurfaces,'' in
  \emph{2015 IEEE International Symposium on Antennas and Propagation \&
  USNC/URSI National Radio Science Meeting}.\hskip 1em plus 0.5em minus
  0.4em\relax IEEE, 2015, pp. 874--875.

\bibitem{achouri2014general}
K.~Achouri, M.~A. Salem, and C.~Caloz, ``General metasurface synthesis based on
  susceptibility tensors,'' \emph{IEEE Trans. Antennas Propag.}, vol.~63,
  no.~7, pp. 2977--2991, July 2015.

\bibitem{7428120}
------, ``Electromagnetic metasurface performing up to four independent wave
  transformations,'' in \emph{2015 IEEE Conference on Antenna Measurements
  Applications (CAMA)}, Nov 2015, pp. 1--3.

\bibitem{Idemen1973}
M.~M. Idemen, \emph{Discontinuities in the Electromagnetic Field}.\hskip 1em
  plus 0.5em minus 0.4em\relax John Wiley \& Sons, 2011.

\bibitem{kuester2003av}
E.~F. Kuester, M.~Mohamed, M.~Piket-May, and C.~Holloway, ``Averaged transition
  conditions for electromagnetic fields at a metafilm,'' \emph{IEEE Trans.
  Antennas Propag.}, vol.~51, no.~10, pp. 2641--2651, Oct 2003.

\end{thebibliography}

\end{document}